# Structural, optical and mechanical properties of $Y_2Ti_2O_7$ single crystal

M. Suganya[a], K. Ganesan[b,c,*], P. Vijayakumar[b], Amirdha Sher Gill[a,*], R. Ramaseshan[b,c] and S. Ganesamoorthy[b,c]

[a]*School of Science & Humanities, Sathyabama Institute of Science and Technology, Chennai, Tamil Nadu, India*
[b]*Materials Science Group, Indira Gandhi Centre for Atomic Research, Kalpakkam, Tamil Nadu, India*
[c]*Homi Bhabha National Institute, IGCAR, Kalpakkam, Tamil Nadu, India*
*Corresponding authors: kganesan@igcar.gov.in; amirdhashergill@gmail.com

## Abstract

We report on the growth of $Y_2Ti_2O_7$ single crystals by optical floating zone technique. X-ray diffraction and Raman spectroscopy studies confirm the structural quality of the crystal. The UV-Vis optical studies reveal that the grown crystals have a high optical transparency with an optical band gap of 3.44 eV. The hardness of $Y_2Ti_2O_7$ single crystal is measured for the first time using nanoindentation. The measured hardness, indentation and bulk modulus are found to be 16.4±0.4, 321.1±6.9 and 243.3±5.2 GPa respectively, which are higher than its polycrystalline counterpart and its constituent metal oxides, $Y_2O_3$ and $TiO_2$.

The pyrochlore rare earth (RE) titanates have attracted much attention of researchers because of their superior structural characteristics and exotic physical properties. These materials possess high melting point, optical nonlinearity, ionic conductivity, low thermal conductivity and good chemical and mechanical stability [1–6]. Yttrium titanate ($Y_2Ti_2O_7$) is one of the important members of the pyrochlore family. It has strong electron–phonon interaction and high oxygen vacancy concentration which results in large defect concentration [1]. The $Y_2Ti_2O_7$ compounds possess high refractive index, high optical band gap and high up-conversion luminescence intensity in the visible wavelength [3,5]. These properties make this material a promising candidate as electrolytes in solid-oxide fuel cells, photo catalyst, high-permittivity dielectrics, host material of optical emission, and thermal barrier coatings [2–5,7].

The $Y_2Ti_2O_7$ compounds also have high tolerance to high energy radiations and they are useful as host material for immobilization of nuclear waste [6]. Moreover, the ultrafine $Y_2Ti_2O_7$ nanoclusters is one of the strengthening additives which provides the mechanical strength and radiation resistance in oxide-dispersion-strengthen alloys. The $Y_2Ti_2O_7$ compounds are known to have high mechanical strength due to the short inter-atomic distances of Ti–O and Y–O bonds, in which Ti-O bond has higher covalent character [8]. Despite a considerable amount of literature is available on the structural, optical and electrochemical properties [2–5,9], the studies on mechanical properties of $Y_2Ti_2O_7$ compounds are limited and still it is an active subject of theoretical and experimental research [8,10–12].

The elastic constants are fundamental properties of materials and are directly related with crystal structure and the nature of chemical bonding within a compound. Here, we provide the existing literature related to the experimental and theoretical calculations on the elastic properties of $Y_2Ti_2O_7$ compound. The calculated elastic and bulk modulus of $Y_2Ti_2O_7$ by different groups are found to be ~ 229 and 183.5 GPa [8]; 447 and 229 GPa [10]; 198.9 and 134.9 GPa [13]; 280 and 181 GPa [14], 309.5 and 208.5 GPa [15] respectively. Danielson et al [16] predicted that the elastic modulus decreases from 263 to 198 GPa with *He* interstitial defects, although elastic constants did not vary significantly. A few experimental works are also reported on nano- or microcrystalline $Y_2Ti_2O_7$ ceramics. He et al. [11] had reported the experimental elastic and bulk modulus of polycrystalline $Y_2Ti_2O_7$ compounds to be in the range of 253-265 and 170-192 GPa, respectively. In addition, Vickers hardness of 11.4 [2] and nanoindentation hardness of 12.1 GPa [11], and bulk modulus of 205 GPa [17] are also reported for polycrystalline ceramics. The only available report on elastic properties of $Y_2Ti_2O_7$ single crystal is studied by resonant ultrasound spectroscopy which provides 263 and 171 GPa corresponding to elastic and bulk modulus, respectively [12]. As discussed above, it is obvious that the elastic properties of $Y_2Ti_2O_7$ is found to vary significantly for different research groups. In addition, a reliable experimental data is also lacking to support the theoretical calculation. These facts motivated us to probe the mechanical properties of $Y_2Ti_2O_7$ single crystal that can help to standardize the nanomechanical properties of $Y_2Ti_2O_7$ and also, can serve as reference data for theoretical calculation.

The single crystals of pyrochlore oxides are generally grown by Czochralski method, flux growth technique and optical floating zone (OFZ) technique [18–20]. Of them, OFZ method has advantages over other growth methods since it is a crucible-free technique. The important factor to grow good quality single crystal by OFZ is to achieve a stable molten zone by controlling the shape of the solid - liquid (S-L) interface. The shape of S-L interface depends on growth parameters such as the lamp power, the speed of seed / feed rotation, growth rate and the applied ambient gas composition and pressure [18–20]. A large deviation from the optimized growth parameters leads to striation, crack and local lattice deformations in the grown crystals.

In the present study, the single crystals of $Y_2Ti_2O_7$ were grown using OFZ technique having four mirrors (FZ-T-4000-H-HR-I-VPO-PC) under continuous flow of air at the flow rate of 0.25 l/min. Prior to growth, polycrystalline $Y_2Ti_2O_7$ powders were synthesized by solid state reactions using high pure $Y_2O_3$ (99.995%) and $TiO_2$ (99.95%) chemicals. Upon confirmation of $Y_2Ti_2O_7$ phase formation by powder XRD, feed and seed rods were prepared using hydrostatic press at 70 MPa. Then, these cylindrical rods were again annealed at 1400 °C for 24 h and used for crystal growth. Due to non-availability of seed, growth was initiated on a polycrystalline rod. The growth parameters were optimized to grow crack free uniform single crystal. At first, the lamp power is continuously increased until the height of the molten zone equals to the diameter of the growing crystal to attain a steady state between the feed rod and growing crystal. Then, the crystal growth is initiated at the growth rate of 15 mm/h along with simultaneous feed – seed counter rotation rate that is varied from 10 to 50 rpm. At low rotation rate of 10 rpm, the molten zone is highly unstable and tends to spill over. With increase in the rotation rate from 10 to 40 rpm, the melt spill over is reduced and also, the molten zone tends to be stable. At higher rotation rate of 40-60 rpm, the molten zone becomes more stable with reduced zone height.

Fig. 1a shows the $Y_2Ti_2O_7$ crystal grown at the growth rate of 15 mm/h with a rotation rate of 50/50 rpm. A lot of cracks are observed in the crystal probably due to high growth rate. Then, the crystal growth is repeated under identical conditions except the growth rate which is reduced to 10 mm/h. Though the melt stability is better at 50 rpm rotation and 10 mm/h growth rate, still the grown crystal has many cracks as can be noticed in the initial growth regime of the crystal shown in Fig. 1b. Hence, the rotation rate is further increased to 60 rpm which results in crack free uniform diameter crystal as shown in Fig. 1b. In fact, the rotation rate helps to modify the shape of the S-L interface and stabilize the molten zone. At low rotation rate, thermocapillary convection is the dominant force in the molten zone and it causes high inhomogeneity in the temperature of the melt with high convex S-L interface. On the other hand, at higher rotation rate, the forced convection becomes dominant over thermocapillary convection and the S-L interface becomes less convex which provides a more stable molten zone [19]. Thus, the present study demonstrates that it is possible to control the molten zone stability and the shape of S-L interface by varying the feed-seed rotation rate during growth of $Y_2Ti_2O_7$ crystals.

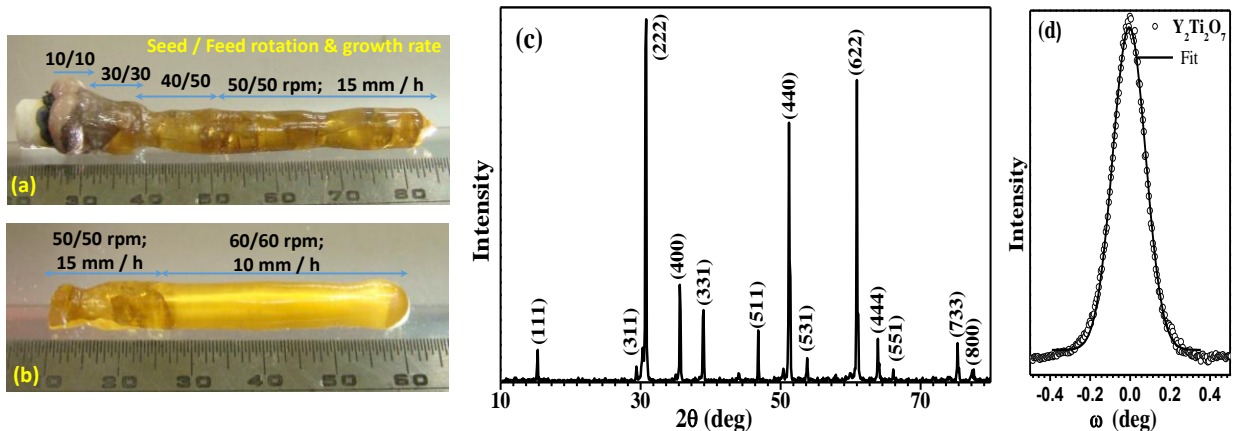

Fig 1. The photograph of the $Y_2Ti_2O_7$ single crystals grown at the growth and rotation rate (a) 15 mm/h and 50 rpm, (b) 10 mm/h and 60 rpm, respectively. (c) X-ray diffraction pattern of powdered $Y_2Ti_2O_7$ single crystal and (d) The rocking curve profile of the crystal, ω scan along <111> orientation.

Subsequent to the successful growth, the phase purity and structural quality of the grown $Y_2Ti_2O_7$ crystals were analyzed by powder X-ray diffraction (PXRD) and rocking curve measurements respectively, using Cu-Kα source (STOE STADI MP-XRD). The UV-Vis-NIR transmission studies were measured using Lambda 35 Perkin Elmer instrument in the wavelength range of 200-1100 nm. The Raman measurements were carried out using Renishaw invia micro-Raman spectrometer with 532 nm diode laser. The hardness, elastic and bulk elastic modulus of the grown crystal were estimated using nanoindentation system (M/s. Anton Paar Ltd.). Indentations were performed using Berkovich diamond indenter with a tip radius of ~ 30 nm with a maximum load of 10 mN, increased at the rate of 20 mN/min, as per the ISO 14577-1 standard [21]. Oliver and Pharr mechanism is followed to extract the hardness and modulus data [22]. The wafer of 1 mm thickness is used for hardness measurements.

Figure 1c shows the PXRD pattern of powdered $Y_2Ti_2O_7$ single crystal. The diffraction pattern matches well with JCPDS card no 00-42-0413. The absence of secondary phases related to Y-Ti-O complex structure confirms the phase purity of the grown $Y_2Ti_2O_7$ crystal. The calculated lattice parameter using Rietveld refinement is found to be 10.086 Å which is consistent with reported value. Further, it is known that the pyrochlore structure undergoes order-disorder transformation into defect fluorite structure when the cation antisite disorder increases. Furthermore, the degree of cation antisite disorder can be estimated from the intensity ratio of (331) and (400) reflections ($I_{331}/I_{400}$) [23]. The estimated $I_{331}/I_{400}$ ratio for the studied $Y_2Ti_2O_7$ crystal is found to be ~ 0.76 which is very close to the theoretically calculated value of 0.71 for an ideal and fully ordered pyrochlore structure [23]. Thus, the XRD pattern confirms that $Y_2Ti_2O_7$ crystal has highly ordered pyrochlore structure with minimal cation antisite disorder. The orientation of the grown crystal is found to be <111>. The full width at the half maximum of the rocking curve is ~ 695 arc-sec which shows the reasonable quality of the crystal (Fig. 1d).

Figure 2 shows the Raman spectrum of the grown $Y_2Ti_2O_7$ single crystal. The spectrum shows six Raman active modes of $A_{1g}$, $E_g$, and $4F_{2g}$ vibrations which are typical for pyrochlore structure. The Raman bands at ~ 218, 308, 450 and 595 cm$^{-1}$ are assigned to $F_{2g}$ phonon vibrations and other bands at ~ 320 and 521 cm$^{-1}$ are due to $E_g$ and $A_{1g}$ mode vibrations, respectively [24]. The highest intensity Raman band at ~ 308 cm$^{-1}$ is due to the combination of $F_{2g}$ and $E_g$ modes and are mostly attributed to O-Y-O bending vibrations. The Raman band at 218 cm$^{-1}$ is assigned to the vibration of oxygen at 8a (O$_2$) located at the center of the Y$_4$O tetrahedral. The $A_{1g}$ mode at ~ 521 cm$^{-1}$ is attributed to the TiO$_6$ octahedra bending vibration. The weak band ~ 450 cm$^{-1}$ is assigned to the $F_{2g}$ mode vibrations [24] and a magnified part of the spectrum is indicated by an arrow mark in the inset of Fig. 2. The other high frequency modes ~ 708 and 890 cm$^{-1}$ are attributed to second order Raman scattering. The weak low frequency mode at ~ 109 cm$^{-1}$ may be due to the infrared active mode that arises due to possible lowering of local symmetry in the lattice [24]. Raman spectrum also confirms that no other impurity phases related to Y-Ti-O compounds is present in these grown Y$_2$Ti$_2$O$_7$ single crystals.

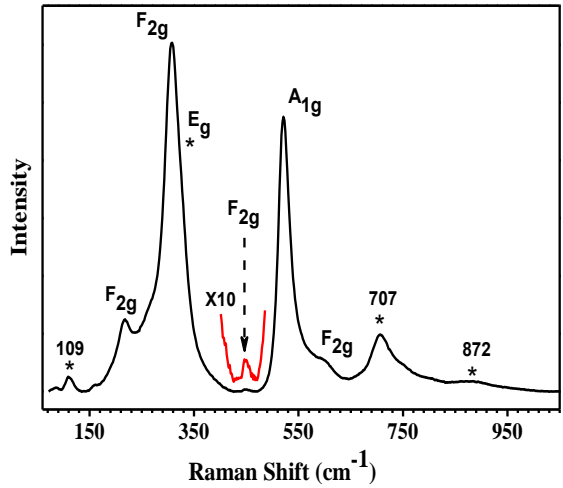

Fig. 2. Raman spectrum of Y$_2$Ti$_2$O$_7$ single crystal at room temperature. A magnified part of the spectrum is given as inset.

The UV-VIS-NIR spectrum of the Y$_2$Ti$_2$O$_7$ single crystal over the wavelength range of 200 – 1100 nm is shown in Fig. 3. The crystal has high transparency of ~ 54 % for 1 mm thickness at 1100 nm and it also confirms the reasonably good quality of the crystal. The optical band gap energy is calculated from Tauc relation,

$$A(h\upsilon - E_g) = (\alpha h\upsilon)^n \quad \ldots\ldots\ldots\ldots\ldots\ldots (1)$$

where, $\alpha$ is the absorption coefficient, A is the proportionality constant, $h\upsilon$ is the photon energy, $E_g$ is optical band gap and n is an integer which defines the type of electronic transition; n=2 for direct bandgap & n=1/2 for indirect bandgap materials. The Tauc plot is given as inset in Fig. 3. The best fit parameters indicate that the Y$_2$Ti$_2$O$_7$ crystal undergoes direct energy transition with band gap of 3.44 eV. This estimated band gap of 3.44 eV is consistent with the reported experimental data for $Y_2Ti_2O_7$ crystal [4,7] while the theoretical calculation always underestimates the band gap of about 2.8 eV due to its limitation [3].

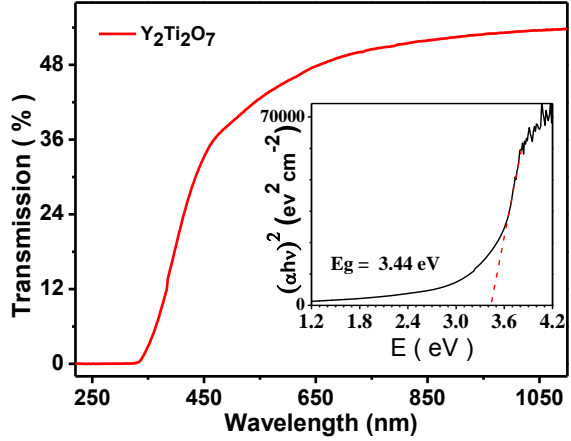

Fig. 3. UV-Vis-NIR optical transmission spectrum of $Y_2Ti_2O_7$ single crystal wafer. The inset shows the Tauc plot with energy gap of 3.44 eV.

Nanoindentation technique is an indispensable tool to measure the local mechanical properties at very precisely on thin films as well as bulk materials. Fig. 4a depicts the typical load – displacement curve measured with a constant load of 10 mN on the $Y_2Ti_2O_7$ single crystal during nanoindentation. An optical micrograph of the indented area is also given in Fig. 4b which shows an identical indentation shape that confirm the uniformity and homogeneity of the indentations. Several indentations are performed in the form of 3X4 matrix on the surface of the single crystal and the results are shown in Table 1.

The hardness ($H_{IT}$) of the single crystal is calculated from the maximum load ($P_{max}$) and the contact depth ($h_c$) using, [22]

$$H_{IT} = \frac{P_{max}}{A_c} \quad \ldots\ldots\ldots\ldots (2)$$

where, $A_c$ is the projected area of the impression; $A_c = 24.5 * h_c^2$, for a perfect Berkovich indenter. The contact depth, $h_c = h_{max} - \varepsilon\beta \frac{P_{max}}{dP/dh}$, where dP/dh is the stiffness, S, which is measured from upper portion of the unloading curve and the S is proportional to reduced elastic modulus, $E_r$, with the relation $S = \frac{2}{\sqrt{\pi}} E_r \sqrt{A_c}$, $h_{max}$ is the probe displacement of the final unloading curve, ε is the strain and β is a dimensionless parameter related to shape of the probe.

Based on the measured $E_r$, the elastic modulus is calculated using the relation, [22]

$$\frac{1}{E_r} = \frac{1-v_m^2}{E_m} + \frac{1-v_i^2}{E_i} \quad \text{------------------------------------------} \quad (3)$$

where $E_m$, $E_i$ and $v_m$, $v_i$ are elastic modulus and Poisson's ratio of specimen and the indenter, respectively. $E_r$ is the combined elastic modulus of the contacting bodies or reduced modulus. The bulk modulus (B) is calculated using elastic modulus (E) by,

$$B = \frac{E}{3(1-2v)} \quad \text{------------------------------------------} \quad (4)$$

where, $v$ is the Poisson ratio which is taken as 0.28 for the calculation.

Table 1 presents the measured $H_{IT}$, E and B using the equations 2, 3 and 4, respectively and also, these values are compared with the existing literature on $Y_2O_3$, $TiO_2$, $Y_2Ti_2O_7$ ceramics. As can be seen from the Table 1, the mechanical properties of the $Y_2Ti_2O_7$ single crystal are higher than its polycrystalline counterpart and its constituent metal oxides, $Y_2O_3$ and $TiO_2$. The hardness of the single crystal is 1.36 times higher; elastic (bulk) modulus is 1.25 (1.59) times higher than polycrystalline samples. Moreover, the grown single crystal displays much higher hardness than the of only available data on $Y_2Ti_2O_7$ crystal reported in literature (10.77 GPa) [25]. Thus, it is evident that the grown crystal shows the highest hardness and modulus values among the available reports in litrature. Further, this study on elastic properties of single crystal help to improve the understanding of interatomic forces in $Y_2Ti_2O_7$ compounds.

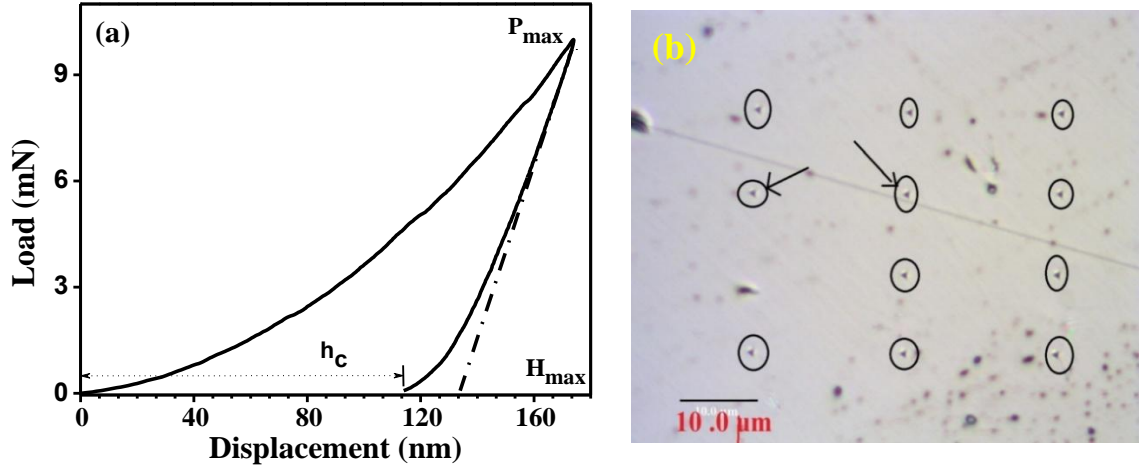

Fig. 4 (a) Nanoindentation mark on the surface of $Y_2Ti_2O_7$ crystal (b) A typical load-displacement curve during nanoindentation on $Y_2Ti_2O_7$ crystal

Table 1. Mechanical properties of $Y_2Ti_2O_7$ single crystal as compared with $Y_2O_3$, $TiO_2$, and $Y_2Ti_2O_7$ polycrystalline ceramics

| Physical properties | $Y_2O_3$ [11] | $TiO_2$ [11] | $Y_2Ti_2O_7$ polycrystalline ceramics | $Y_2Ti_2O_7$ crystal | $Y_2Ti_2O_7$ Crystal (This work) |
|---|---|---|---|---|---|
| Hardness ($H$) GPa | 6.9-9.1 | 11.0 | 12.1 [11] [a] <br> 11.4 [2] [b] | 10.77 [25] | 16.4+0.4 |
| Elastic modulus ($E$) GPa | 188.0 | 270.0 | 253 [11] [c] | 263 [12] [e] <br> 199-447 [f] | 321.1+6.9 |
| Bulk modulus ($B$) GPa | 158.0 | 152.5 | 170 [11] [c] <br> 205 [17] [d] | 171 [12] [e] <br> 135-229 [f] | 243.3±5.2 |

[a] Nanoindentation, [b] Vickers hardness, [c] Resonance method, [d] High pressure X-ray diffraction,
[e] Resonance ultrasound spectroscopy and [f] A discrete value in this range is calculated by different theory groups [8,10,13–16]

In conclusion, $Y_2Ti_2O_7$ single crystals are successfully grown by optimizing the seed / feed rotation growth rate using optical floating zone technique. The X-ray diffraction and Raman spectroscopic studies establish the structural quality of the crystals. Further, the high optical transparency with optical band gap of 3.44 eV is also ascertain the quality of the crystal. For the first time, the hardness, elastic and bulk modulus of $Y_2Ti_2O_7$ single crystals are measured using nanoindentation and the values are found to be higher as compared to the available reports in litrature. Further, the obtained mechanical parameters of $Y_2Ti_2O_7$ single crystals can serve as reference data for theoretical calculation and to its nano- and microcrystalline counterparts.

The authors, M.S and A.S.G., gratefully acknowledge University Grants Commission (UGC) Indore, India under UGC-DAE-CSR (CSR-KN/CRS-87/2016-17/1128) project scheme for financial assistance. Also, M.S and A.S.G thank Dr. N.V. Chandrasekar, Scientist in-charge, UGC-DAE-CSR, Kalpakkam node for his constant support and encouragement. The authors also thank Dr. R.M. Sarguna and Mrs. Sunitha Rajakumari of Materials Science Group, IGCAR, Kalpakkam for their support in X-ray diffraction and crystal polishing, respectively.

Declaration of interests

The authors declare that they have no known competing financial interests or personal relationships that could have appeared to influence the work reported in this paper.